\documentclass[12pt]{article}
\usepackage{latexsym}
\usepackage{amssymb}
\usepackage[dvips]{graphicx}
\usepackage{epsfig}
\usepackage{rotating}

\usepackage{color}
\usepackage{cancel}

\usepackage{amsfonts}
\usepackage{amsmath}

\newcommand{\bea}{\begin{eqnarray}}
\newcommand{\eea}{\end{eqnarray}}
\newcommand{\beq}{\begin{equation}}
\newcommand{\eeq}{\end{equation}}

\newcommand{\tr}{{\rm tr}}
\newcommand{\qed}{\hfill $\square$}
\newcommand{\proof}{{\bf Proof.\;}}

\newcommand{\R}{\mathbb{R}}

\newcommand{\cM}{\mathcal{M}}
\newcommand{\cJ}{\mathcal{J}}

\newcommand{\cP}{\mathcal{P}}
\newcommand{\mSn}{\mathcal{S}_{N}}
\newcommand{\mSi}{\mathcal{S}_{\infty}}

\newcommand{\ket}{\rangle}

\newcommand{\Bra}{\Big{\langle}}
\newcommand{\Ket}{\Big{\rangle}}

\newtheorem{proposition}{Proposition}
\newtheorem{theorem}{Theorem}

\makeatletter

\begin{document}

\begin{titlepage}

\begin{flushright}
pi-mathph-330\\
ICMPA-MPA/2013/005\\
\end{flushright}

\begin{center}

{\Large\bf The finite and large-$N$ behaviors \\
\medskip of independent-value matrix models}

\bigskip

Joseph Ben Geloun$^{a,c,*}$
 and John R. Klauder$^{b,\dag}$

\bigskip

$^a${\em Perimeter Institute for Theoretical Physics, 31 Caroline
St N} \\
{\em ON N2L 2Y5, Waterloo, ON, Canada} \\
\medskip
$^{b}${\em Department of Physics and
Department of Mathematics}\\
{\em University of Florida, Gainesville, FL 32611-8440}\\
\medskip
$^{c}${\em International Chair in Mathematical Physics
and Applications}\\
{\em ICMPA--UNESCO Chair, 072 B.P. 50  Cotonou, Republic of Benin} \\
\medskip
E-mails:  $^{*}$jbengeloun@perimeterinstitute.ca,
\quad $^{\dag}$klauder@phys.ufl.edu

\begin{abstract}
We investigate the finite and large $N$ behaviors of independent-value
$O(N)$-invariant matrix models. These are models defined with matrix-type fields
and with no gradient term in their action. They are generically nonrenormalizable but can be handled
by nonperturbative techniques. We find that the functional of any $O(N)$ matrix  trace invariant
may be expressed
 in terms of an $O(N)$-invariant measure. Based on
 this result, we prove that, in the limit that
 all interaction coupling constants go to zero, any interacting theory is continuously connected to a
pseudo-free theory. This theory differs radically from the familiar free theory consisting
in putting the coupling constants to zero in the initial action. The proof is given
for generic finite-size matrix models, whereas, in the limiting case $N\rightarrow\infty$, we succeed
in showing this behavior for restricted types of actions using a particular scaling of the parameters.
\end{abstract}

\end{center}

\noindent Pacs numbers: 11.15.Pg, 11.10.Jj, 04.60.Nc


\vspace{0.5cm}
\begin{center}
\today
\end{center}
\end{titlepage}

\setcounter{footnote}{0}

\section{Introduction}

Large-$N$ expansion techniques are useful in the study of systems with an infinite number of degrees of freedom in quantum field theory \cite{tooft}. In the case of many-body systems, this method has led to the well known
Hartree-Fock-type approximations  \cite{revhf}. The approach also proves to be crucial when dealing with nonrenormalizable models in order to localize the main source of divergences and to find a way to extract these,
for instance, in an infinite lattice regularization scheme
\cite{Klauder:1975kd,Klauder:1975kc,Klauder:2012uf,Klauder:2011gv}.
The models under discussion here have the property of infinite divisibility (defined below), which, as is well known \cite{luka}, exhibits both Gaussian and Poisson behavior. The former (Gaussian) behavior typically applies to free models, while  interacting models typically involve the latter (Poisson) behavior.

The specific kind of problems we will be dealing with in this work are of the nonrenorma- lizable type.
Such models can occur more often that one may think.
Indeed, on one hand,  the  criteria for a theory to be renormalizable are specific and express a fine balance
between several ingredients of the theory \cite{Rivasseau:1991ub}.\footnote{The main ingredients of a theory is its space dimension,
its propagator, the type and valence of its vertices. Renormalizability, at least at the perturbative
level, expresses the self-replicability or stability  or locality principle of the model under change
of energy scales. It is captured by a power-counting theorem and a locality principle
issued, for instance, from a specific multi-scale analysis \cite{Rivasseau:1991ub}.} On the other hand,
one agrees with the following fact: Consider a scalar field theory in a Euclidean $D$-dimensional space characterized by a functional integral  (in standard notation  using a time-ordered product)
\beq
Z(J)= K \int [D\phi]\; e^{i \int J (x)\phi(x)d^Dx -\int  \left(\frac12 \partial^\mu\phi(x) \partial_\mu\phi(x)+ \frac12 m^2 \phi(x)^2+
\lambda V(\phi(x))  \right) d^Dx}\,,
\label{zpart}
\eeq
where $K$ is a normalization, $J$ is a real source field---we have chosen an imaginary multiplier to make closer contact with the language of characteristic functions, i.e., Fourier transforms of probability distributions---$m$ is the mass, and $V$ is the interaction with coupling $\lambda$.
If we {\it discard} the gradient term $\partial^\mu\phi(x) \partial_\mu\phi(x)$ altogether, we reach a theory with propagator $1/m^2$ hence without ultraviolet (UV) momentum damping, which will lead to arbitrarily many divergences in a perturbative study of the expression. When there is no gradient term in the action,
the field is statistically independent at each point $x$ of spacetime
and, consequently, no excitation can spread from one point to another.
Following earlier  practice \cite{Klauder:1975kd}, we call these models {\it independent-value models}.
Such models with no damping in momentum,
would be generically nonrenormalizable at the perturbative level.

We should emphasize that models with no gradient in the kinetic term are not normally
encountered in relativistic quantum field theory because of the kinematic behavior
they possess.  However, there are several models in statistical mechanics, especially in
matrix models \cite{difrance}, for which the kinetic term is exactly of the modified kind.
Thus such models are certainly worthwhile to discuss.

Closer to the interest of the present work has been an investigation of the large-$N$ behavior
of independent-value vector models \cite{Klauder:1975kd,Klauder:1975kc}.
In these contributions, $O(N)$-invariant Euclidean actions defined over vectors $\phi=(\phi_k)_{k=1}^{N}$ have been studied at finite $N$ and in the limit $N=\infty$.  Using nonperturbative
techniques, the authors showed, both at finite and infinite $N$,  that
the solutions of any interacting theory do {\it not} reduce to those of the free theory in the
limit where the interaction coupling constant goes to zero. Moreover, using a nontraditional  asymptotic dependence
of $N$ on the several parameters of the interacting theories,  the limit $N\to \infty$ does {\it not} lead to a conventional free solution. This new solution, which arose by continuity when the interaction is reduced to zero, is called {\it the pseudo-free theory}, and when it is different from the free theory, it must be considered as
the theory about which to expand the interacting theory. As a result, the existence and use of the pseudo-free theory is highly important in remedying certain pathological features of some otherwise puzzling models \cite{Klauder:2012uf}.

In this paper, we extend the results obtained for vector models to
$O(N)$-invariant  matrix models. Both finite and infinite size $N\times N$
matrices are addressed. The formulation is focused on symmetric matrices but,
without much additional work, our results can be extended to any real or complex square matrices.
We prove as well that the interacting matrix theories are continuously connected to a suitable pseudo-free
matrix theory. Our solution rests on the existence of a model-dependent, invariant measure
which needs to be determined to fully specify the solution. Indeed, we can find the explicit and nontrivial expression for this invariant measure
in  the case of a finite-size matrix, whereas for the infinite-size matrix case, we succeed
to find a nontrivial measure under some restrictions regarding the type of invariant models
when $N=\infty$.

The paper is organized as follows. The next section reviews the previous
work \cite{Klauder:1975kd} which is our guiding thread from the vector case toward the
matrix case, which is studied in Section \ref{sect:mat}. In the last paragraph of that section, we comment
on how our results extend to real matrices (the step from real matrices to complex
ones would be straightforward). Theorem \ref{theo1} is one of our main results
in this paper. Given a model, Section \ref{sect:app} investigates the explicit invariant measure as stated in Theorem \ref{theo1}
which is another important result of this work. In Section \ref{sect:concl} we summarize
our results as well as include comments on future projects.

\section{Independent-value vector models: a review}
\label{sect:rev}

This section undertakes a review of the results obtained
for vectors and is largely based on \cite{Klauder:1975kd}.

\medskip

\noindent{\bf Finite-component vectors.}
Consider  a real vector field $\phi(x)=\{\phi_k(x), k=1,\dots, N\}$,
$x \in \R^n$, with $n \geq 1$ and  a model described by the $O(N)$-invariant Euclidean action built from $\phi\,$:
\bea
S_\lambda[\phi]= \int_{\R^n} \left( \frac12 m^2 |\phi(x)|^2+
\lambda V(|\phi(x)|^2) \right) dx\,,
\label{act:iid}
\eea
where $ |\phi(x)|^2 \equiv\sum_{k}\phi_k(x)^2$. $S_{\lambda=0}$
corresponds to the free theory.
We call these models independent-value vector models, which have been
studied in \cite{Klauder:1975kd, Klauder:1975kc}.
Such models with no damping in momentum,
would be generically nonrenormalizable at the perturbative level
as already discussed in the first section.

The functional integral with a source $J(x)=\{J_k(x), k=1,\dots, N\}$ in the Euclidean formulation,  related to the action $S$, Eq. \eqref{act:iid}, is given by
\bea
Z(J, \lambda) = {\cal N}\int [\prod_{x;k} d\phi_k(x)] \; e^{i \int (J \cdot \phi)(x)dx  - S_{\lambda}[\phi]}\,.
\label{part}
\eea
Due to the lack of gradients and the rotation invariance under $\phi \to O \phi$,
where $O \in O(N)$, we can write
\bea
Z(J,\lambda) = e^{-\int L_\lambda(|J(x)|)dx}\,,
\label{zj}
\eea
where $L_\lambda$ is an  $O(N)$-invariant function of the source $J$. From its construction, the characteristic function $Z(J, \lambda)$ is infinitely divisible,  i.e., $Z(J, \lambda)^{1/m}$ is also a characteristic function for all positive integers $m$ \cite{luka}. We next exploit that very fact.

Let us assume that $J = P \chi_{\Delta}(x)$, where we
separate the vector part $P=\{P_k\}\in \R^N$ and
the background space via $\chi_{\Delta}(x)$ the indicator function of
 $\Delta_*$ a compact subset  of $\R^n$ with volume $\Delta$;
precisely, if $x \in \Delta_*$, then $\chi_{\Delta}(x)=1$, otherwise $\chi_{\Delta}(x)=0$. Then, we have
\bea
Z(J,\lambda)  =  e^{-\Delta L_\lambda(|P|)} = \int_{\R^N} \cos (P \cdot u) \, d\mu_{\Delta,L_\lambda}(u)\,,
\label{zj2i}
\eea
for some $O(N)$-invariant probability measure $d\mu_{\Delta,L}(u)$ on $\R^N$. Hence, it is straightforward to obtain
\beq
L_\lambda(|P|) = \lim_{\Delta \to 0} \Delta^{-1}\left( \int_{\R^N} [1-\cos (P \cdot u)] \, d\mu_{\Delta,L_\lambda}(u)\right) .
\eeq
The most general form for such a limit is given by \cite{luka} as
\beq
L_\lambda(|P|) = a |P|^2 + \int_{|u|>0} [1-\cos (P \cdot u)] \, d\sigma_\lambda(u) \,,
\label{Llam}
\eeq
where $a\geq 0$ and
$d\sigma_\lambda$ is an $O(N)$-invariant nonnegative measure subject to
the condition that
\bea
\int_{|u|>0} [|u|^2/ (1+ |u|^2)]  d\sigma_\lambda(u)  < \infty\,.
\eea
Interestingly, we note that:

- the Gaussian cases, $a >0$ and $\sigma_\lambda\equiv 0$, yields all free theories with different masses;

- the Poisson cases, $a\equiv 0$ and $\sigma_\lambda \neq 0$, cover all
interacting theories (i.e., {\it non}free theories). 

While it is possible to consider both terms being nonzero, local powers of the Gaussian field or of the Poisson field are made in very different ways, as we shall soon see, and thus, for quantum field applications where local products are important, it is necessary to consider the Gaussian and Poisson cases separately.  Finally, for the Poisson cases, given an interacting theory with coupling
$\lambda$, the remaining task is
to find the measure $\sigma_\lambda$ associated with it.

\medskip

\noindent{\bf Infinite-component vectors.}
Assuming that we are dealing with an infinite component vector
field $\phi(x)=\{\phi_k(x)\}_{k=1}^{\infty}$, the above analysis
extends in the following way. Eq. \eqref{zj} holds still with
$L(|J|)$ an even $O(\infty)$-invariant function. Using again
particular fields $J= P\chi_{\Delta}$, we can formulate \eqref{zj2i} as
\bea
\int \cos(P \cdot u) d\mu_{\Delta, L_\lambda}(u)
 = \int_0^\infty e^{- b \,|P|^2} d\mu_{\Delta, L_\lambda}(b)\,,
\label{mcosN}
\eea
where use has been made of the fact that every characteristic
function with $O(\infty)$ invariance is a convex combination
of Gaussians \cite{schoen}.

We derive from \eqref{mcosN} that
\bea
L_\lambda(|P|) = \lim_{\Delta \to 0} \Delta^{-1}\left( \int_0^\infty [1-e^{-  b \,|P|^2} ] \, d\mu_{\Delta,L_\lambda}(b)\right),
\eea
with the most general form  given by
\bea
L_\lambda(|P|) = a |P|^2 + \int_{0}^{\infty}  [1-e^{- b \,|P|^2} ] \, d\sigma_\lambda(b) \,,
\eea
where $a\geq 0$, $d\sigma_\lambda$ is a nonnegative measure on $(0,\infty)$
obeying the condition
\bea
\int_{0}^{\infty} [b/ (1+ b)]  d\sigma_\lambda(b)  < \infty\,.
\label{bbound}
\eea
The same remarks associated with free theories
characterized by $a>0$ and $\sigma_\lambda\equiv 0$
hold here. These provide $N=\infty$ free solutions for vector models.
Meanwhile, $a\equiv 0$ and $\sigma_\lambda\neq 0$ covers
interacting theories.

In \cite{Klauder:1975kd}, as an illustration, for the model \eqref{act:iid} and
under suitable conditions, one can establish that
\bea
&& N<\infty\,, \qquad d\sigma_\lambda(u) =C\, e^{-\frac12 m^2 u^2 - \lambda V(u^2)} \frac{1}{|\vec u|^{N}} \,d\vec u\,,\crcr
&& N=\infty\,, \qquad d\sigma_\lambda(b) =C\, e^{-\frac12 m^2 b - \lambda V(b)} \frac{1}{|b|} \,db\,,
\eea
for a suitable value of $C$ in each case.
\section{Independent-value matrix models}
\label{sect:mat}

We now investigate analogs of the statements in Section \ref{sect:rev} for real matrix models.
The simple way to address this is to consider real symmetric matrices
$M(x)=\{ M_{ab}(x)=M_{ba}(x)\}_{a,b=1,\dots,N}$, where $x\in \R^n$.
There exists an extension of the following discussion for non-symmetric matrices and even
complex ones, however the basic ideas would remain  the same. We will
come back to this point later on.

Consider now an $O(N)$-invariant action which can be written as
\bea
S_{\vec \lambda}[M] = \int_{\R^n} \left[\, \frac12\mu^2\, \tr [M^2(x)] +
 V_{\vec \lambda}\Big(\{\tr [(M(x))^{2p}]\}_{p} \Big) \right]dx\,,
\label{actmatrix}
\eea
where $\mu$ is the mass parameter and $V_{\vec \lambda}$ is an $O(N)$-invariant function of some of the
invariants $\tr [(M(x))^{2p}]$, for $p$ an integer such that $2\leq p \leq p_{\max}$,
and we can consider $p_{\max}$ finite or
infinite. Furthermore,
$\vec \lambda=(\lambda_2,\dots, \lambda_{q})$
which simply collects all coupling constants depending on the number
of terms involved in $V_{\lambda}$. A typical situation would be
to consider an interaction  of the form
\beq
 V_{\vec \lambda}\Big(\{\tr [M^{2p}]\}_{p} \Big)
 = \sum_{p=2}^{p_{\max}} \lambda_{p} V_{p}(\tr(M^{2p}))\qquad
\eeq
where $\vec\lambda=(\lambda_2,\dots, \lambda_{p_{\max}})$. Note that the expression $\det[M(x)]$ and its
powers $\det[M^p(x)]=(\det[M(x)])^p$ are also
$O(N)$-invariant functions that could be considered for the interaction term. For reasons of simplicity, we do not include such terms.

One notices that again the model is independently  distributed
at each spacetime point $x$. Hence, we call these independent-value
matrix models and write the corresponding  characteristic functional integral for such a model as
\beq
Z(J,\vec\lambda) =K\int \Big[\prod_{x;ab} dM_{ab}(x)\Big]\,
e^{i\int [\tr (J(x)M(x)]dx - S_{\vec \lambda} [M]} \,,
\label{matmo}
\eeq
where $K$ is a normalization factor and $J$ can be chosen as a symmetric matrix.

The following statement holds.
\begin{proposition}\label{prop:mat}
\bea
Z(J,\vec\lambda) = e^{-\int L_{\vec \lambda}[\,\{\tr (J(x)^{2p})\}_{p}\,]dx}\,.
\eea
where $L_{\vec \lambda}$ is a $O(N)$-invariant function of all possible invariants
in  $\{\tr (J(x)^{2p})\}_{p}$.
\end{proposition}
\proof We start by slicing $\R^n$ in $\{\Delta_\ell\}_\ell$ where each $\Delta_\ell \subset \R^n$ has a fixed finite volume $\Delta$ and
 an indicator function $\chi_{\Delta_\ell}$. The index $\ell$ depends
on the slicing but should typically run over an  infinite discrete
set because $\R^n$ is non-compact. Then, we write the matrix field $M(x)= \sum_\ell M_\ell(x)$, where
$M_{\ell}=\cM_{\ell}\,\chi_{\Delta_\ell}$, and $\cM_{\ell}= (\cM_{\ell; ab})$
is a symmetric matrix. Similarly we introduce $J(x) = \sum_{\ell}J_\ell(x)$ and $J_\ell =\cJ_\ell\, \chi_{\Delta_\ell}$, where $\cJ_\ell$ is a symmetric matrix as well.
The functional integral can be re-expressed as
\beq
Z(J,\vec\lambda) = \lim_{\Delta\to 0}\int [\prod_{\ell; a,b}d\cM_{\ell;ab}]
e^{\Delta [\, i\,\tr (\cJ_\ell \cM_\ell) - S_{\ell;\lambda}[\cM]]}
 = \lim_{\Delta\to 0}\prod_{\ell}\int [\prod_{a,b}d\cM_{ab}]
e^{\Delta [\, i\,\tr (\cJ_\ell \cM) - S_{\ell;\lambda}[\cM]]}\, .
\eeq
To rely on a well-defined continuum limit with $O(N)$-invariance
of the matrix $\cJ$ (as $Z$ in \eqref{matmo} could be), the infinite product must be of the form
\bea
Z(J,\vec\lambda) &=&
\lim_{\Delta\to 0}\prod_{\ell} [1 - \Delta C_{\Delta,\ell, \vec \lambda}[ \{\tr (\cJ^{2p})\}_{p}] ]
 =
 \lim_{\Delta \to 0 } e^{- \Delta\sum_\ell L_{\ell, \vec \lambda}( \{\tr (\cJ^{2p})\}_{p})}  \crcr
&=&  e^{-\int_{\R^n} [\sum_\ell \chi_{\Delta_\ell}L_{\ell, \vec \lambda}( \{\tr (\cJ^{2p})\}_{p})]dx }
= e^{- \int L_{\vec \lambda}(\{\tr (J(x)^{2p})\}_{p} )  dx}\,,
\eea
where $C_{\Delta,\ell, \vec \lambda}[ \{\tr (\cJ^{2p})\}_{p}]$ is so defined such that $ L_{\ell, \vec \lambda}( \{\tr (\cJ^{2p})\}_{p})$ is independent of $\Delta$.

\qed

Assuming that $J = \cJ \chi_{\Delta_*}$, $\Delta_*\subset \R^n$
with volume $\Delta$, we simply rewrite
\bea
Z(J,\vec\lambda) = e^{- \Delta L_{\vec\lambda}( \{\tr (\cJ^{2p})\}_{p})}\,.
\eea
On the other hand, a direct reduction from \eqref{matmo} leads to
\beq
Z(J,\vec\lambda) = e^{- \Delta L_{\vec\lambda}( \{\tr (\cJ^{2p})\}_{p})}
 = \int_{\mathcal S_N}\, e^{i\; \tr (\cJ U)} d\mu_{\Delta, \vec \lambda} (U)\,,
\label{zej}
\eeq
where $U$ is a real symmetric matrix and $d\mu_{\Delta, \vec \lambda} (U)$
is an invariant probability measure on the space $\mathcal S_N$ of real $N \times N$  symmetric matrices.
Note that we do not need to integrate over the entire space of matrices
$M_{N}(\R)$, because $\cJ$ is symmetric and the trace will
necessarily yield a reduction on $\mathcal S_N$ provided that
the measure $\nu_{(\cdot)}$ on $M_{N}(\R)$ is factorized so that
$d\nu_{(\cdot)} = d\nu'_{(\cdot)} d\nu''_{(\cdot)}$,
where $d\nu''$ is a probability measure on the space of
anti-symmetric matrices.

\begin{theorem}\label{theo1}
Let $\cJ$ be a symmetric matrix of order $N$ (finite or infinite) and
let $d\mu_{\Delta, \vec \lambda}$ be an invariant probability measure
on the set of real symmetric matrices of order $N$ such that
the following limit converges
\beq
\lim_{\Delta\to 0} \Delta^{-1}\int_{\tr (U^2)>\Delta} f(U) d\mu_{\Delta, \vec \lambda}(U) = \int_{\tr (U^2)>0} f(U)  d\sigma_{\vec \lambda}(U)\,,
\label{hyp}
\eeq
for suitable functions $f$ on $\mathcal S_N$, which are independent of $\Delta$.
Then any  invariant $O(N)$-invariant function built over $\cJ$
with some couplings $\vec \lambda=\{\lambda_i\}_i$ satisfies
\beq
L_{N,\vec\lambda}(\{\tr [\cJ^{2p}]\}_p)
 = a \,\tr [\cJ^2] + \int_{\tr (U^2)>0} (1- e^{i\,\tr (\cJ U)}) d\sigma_{\vec \lambda}(U) \,,
\label{unimat}
\eeq
where $a\geq 0$ and $d\sigma_{\vec\lambda}(U)$ is a nonnegative
invariant measure over $\mathcal S_N$ such that
\beq
\int_{\tr(U^2)>0} \tr(U^2)/(1+\tr(U^2)) d\sigma_{\vec\lambda}(U) < \infty\,.
\label{matcond}
\eeq
In the case of infinite size symmetric matrices $\cJ$,
 or $N\to\infty$ limit, the corresponding  $O(\infty)$-invariant function reads
\beq
L_{\infty,\vec\lambda}(\{\tr [\cJ^{2p}]\}_p) = a \,\tr [\cJ^2] + \int_0^{\infty} (1- e^{-b\; \tr [\cJ^2]}) d\sigma_{\vec \lambda}(b) \,,
\label{larunimat}
\eeq
where $a\geq 0$ and  $d\sigma_{\vec\lambda}(b)$ is a non negative
measure over $\R$ such that
\beq
\int_0^{\infty} b/(1+b)\, d\sigma_{\vec\lambda}(b) < \infty\,.
\label{larmatcond}
\eeq
\end{theorem}
\proof From \eqref{zej}, we have
\beq
L_{N,\vec\lambda}(\{\tr[\cJ^{2p}]\}_p)
= \lim_{\Delta \to 0} \Delta^{-1}\Big(\int_{\mathcal S_N}
(1- e^{i\; \tr (\cJ U)}) d\mu_{\Delta, \vec \lambda} (U)\Big) \,.
\label{lag}
\eeq
\noindent{\bf Case $N$ finite.} We start by decomposing $\int_{\mathcal S_N}$ as
$\int_{\tr(U^2)> \Delta} + \int_{\tr(U^2)\leq \Delta}$ and
evaluate
\beq
 A_{\leq \Delta} =
\int_{\tr(U^2) \leq \Delta}
(1- e^{i\; \tr (\cJ U)}) d\mu_{\Delta, \vec \lambda} (U) =
-\sum_{k=1}^{\infty} \frac{i^{2k}}{(2k)!}\int_{\tr(\widetilde{U}^2) \leq \Delta}
[\tr (j \widetilde{U})]^{2k}\, d\mu_{\Delta, \vec \lambda} (\widetilde{U})\,,
\label{exps}
\eeq
where we used the fact that the measure is invariant in
order to cancel all odd powers in $\tr (\cJ U)$, then we diagonalize $\cJ = OjO^t$
(with $O\in O(N)$ and $j$ diagonal) and
introduce $\widetilde{U}=O^tUO$ such that
$d\mu_{\Delta, \vec \lambda} (\widetilde{U})=d\mu_{\Delta, \vec \lambda} (U)$. We concentrate on the first two terms of the series, which are
\bea
\tr (j \widetilde{U}) = \sum_{a=1}^{N}j_{aa} \widetilde{U}_{aa}\,,
\quad
[\tr (j \widetilde{U})]^{2}= \sum_{a=1}^{N} j^{2}_{aa} \widetilde{U}^{2}_{aa}  +  \sum_{a\neq b}j_{aa} \widetilde{U}_{aa}j_{bb} \widetilde{U}_{bb}\,.
\eea
The first term and the very last sum $\sum_{a\neq b}$ including cross terms vanish because of
the  $O(N)$-symmetry. The next relevant term of the series is
\bea
\tr [(j\widetilde{U})^{4}] = \sum_{a} j^{4}_{aa} \widetilde{U}^{4}_{aa} +
6 \sum_{a\neq b} j^{2}_{aa} \widetilde{U}^{2}_{aa}j^{2}_{bb} \widetilde{U}^{2}_{bb} + \dots \,,
\eea
where the dots include cross terms of odd power which should vanish
by symmetry as well. Notice  that the $2p$-order term
$[\tr (j \widetilde{U})]^{2p}$ in the expansion,
cumulates to a power of $2p$ in the variable $\widetilde{U}_{aa}$.
For simplicity, we write $[\tr (j \widetilde{U})]^{2p} =
O(\widetilde{U}^{2p})$.
 We then re-express \eqref{exps} as
\bea
A_{\leq \Delta} &=&
\frac{1}{2} \sum_{a=1}^{N} j^2_{aa}
\int_{\tr(\widetilde{U}^2) \leq \Delta } \widetilde{U}_{aa}^{2} \, d\mu_{\Delta, \vec \lambda} (\widetilde{U})\crcr
&&
- \frac{i^4}{4!} \left(\sum_{a=1}^{N} j^4_{aa}
\int_{\tr(\widetilde{U}^2) \leq \Delta } \widetilde{U}_{aa}^{4} \, d\mu_{\Delta, \vec \lambda} (\widetilde{U})
 + \sum_{a\neq b} j^2_{aa}j^2_{bb}
\int_{\tr(\widetilde{U}^2) \leq \Delta } \widetilde{U}_{aa}^{2} \widetilde{U}_{bb}^{2} \, d\mu_{\Delta, \vec \lambda} (\widetilde{U}) \right)  \crcr
&&
+ \int_{\tr(\widetilde{U}^2) \leq \Delta } O (\widetilde{U}^{6}) \, d\mu_{\Delta, \vec \lambda} (\widetilde{U})\crcr
 &=&
\frac{\Delta}{2} \sum_{a=1}^{N} j^2_{aa}
\int_{\tr(U'^2) \leq 1 } \frac{\tr[U'^{2}]}{N} \, d\mu_{\Delta, \vec \lambda} (\sqrt{\Delta}U')
+ \Delta^2 \int_{\tr(U'^2) \leq 1} O (U'^{4}) \, d\mu_{\Delta, \vec \lambda} (\sqrt{\Delta}U')\crcr
&&+ \Delta^3\int_{\tr({U'}^2) \leq 1 } O ({U'}^{6}) \, d\mu_{\Delta, \vec \lambda} (\sqrt{\Delta}{U'})\;,
\label{uprim}
\eea
where we used the fact that $\langle \widetilde{U}^2_{aa}\rangle
= \langle \widetilde{U}^2\rangle/N$ and then changed variables so that $U' = \widetilde{U}/\sqrt{\Delta}$.
We now choose $a=\langle \tr[U'^2]\rangle/N\geq 0$.
Note that the measure should be chosen such that $a$
neither depends on  $N$ nor on $\Delta$.  We can now substitute \eqref{uprim} in \eqref{lag}, then using \eqref{hyp}, it is direct to recover \eqref{unimat}. The condition \eqref{matcond} ensures the finiteness
of the integral, namely:   %part $\int_{\tr (U^2)>0}$:
\beq
\int_{\tr (U^2)>0} (1- e^{i \tr (\cJ U)})d\sigma_{\vec\lambda} <\infty \;
\Leftrightarrow \;
\int_{\tr (U^2)>0} \tr(U^2)/(1+\tr(U^2)) d\sigma_{\vec\lambda}(U) < \infty\,.
\eeq

\medskip

\noindent{\bf Case $N=\infty$.}
Let us turn now to the case $N=\infty$ and show \eqref{larunimat}. The expression \eqref{lag} is again valid.
We use the eigenvalue
decomposition of $\cJ=O j O^t$, change variable $U\to \widetilde{U}=OU O^t$ and write
$\tr (j \widetilde U) = \sum_{a} j_{aa} \widetilde U_{aa}$
so that, introducing $r^2=\tr(\widetilde U ^2)= \sum_{ab} \widetilde U_{ab}^2$, we set
\beq
\widetilde U_{11} = r \cos\theta_1 \,,
\quad \text{ and  for } a\geq 2\,,\quad \widetilde
U_{aa} = r \Big(\prod_{l=1}^{a-1}\sin\theta_l\Big) \cos\theta_{a}\,,
\label{diagonal}
\eeq
and, with $N^*= N(N+1)/2$, we re-express \eqref{zej} in spherical coordinates as
\bea
&&\lim_{N \to \infty}
\int_{\mathcal S_N} e^{i\, \tr (\cJ U)}) d\mu_{\Delta, N,\vec \lambda} (U) =\crcr
&&\lim_{N \to \infty} \int_{\mathcal S_N} e^{i r\, \sum_{a}j_{aa} \Big(\prod_{l=1}^{a-1}\sin\theta_l\Big) \cos\theta_{a}} w_{\Delta,N, \vec \lambda} (r)  r^{N^*-1} dr \Big[\prod_{l=1}^{N} (\sin\theta_l)^{N^*-(l+1)}d\theta_l \Big] d\Omega_{N^*-(N+2)} \crcr
&& \label{zejN}
\eea
At large $N$, $N^*-(l+1)\geq N^2 -(N+1) \sim N^2$ is large too, for all $l=1,\dots,N$.
Again, using a steepest-descent technique, one obtains, for each $\theta_l$ variable,  a saddle point $\theta_{l,*}=\pi/2$. Then, we change variable
$\Theta_{l}= \theta_l -\theta_{l,*}$ such that
\beq
 \lim_{N \to \infty} \int_{[-\pi/2,\pi/2]^{N}} e^{i r\; \sum_{a}j_{aa} \Theta_{a}} e^{-\sum_l \frac{(N^*-(l+1))}{2} \Theta_l^2 }
\Big[\prod_l d\Theta_l \Big]
 = \lim_{N \to \infty} k_{N}  e^{- \sum_{a} \frac{1}{2(N^*-(a+1))} j_{aa}^2r^2}
\eeq
for some constant $k_N$. Substituting this result in \eqref{zejN}
and considering that the integration with  $d\Omega_{N^*-(N+2)}$
contributes
at most to an overall factor, we get at large $N$
\bea
\int_{\mathcal S_N} e^{i\; \tr (\cJ U)}) d\mu_{\Delta, N, \vec \lambda} (U)
\simeq
K_{N} \int_0^{\infty}  e^{- \sum_{a} \frac{1}{2(N^*-(a+1))} j_{aa}^2r^2}
w_{\Delta,N, \vec \lambda} (r)  r^{N^*-1} dr \,,
\eea
where $K_N$ includes all constants depending on $N$.
We change variable as $r\to  \sqrt{2bN^*}$ such that, after taking the limit
$N\to \infty$, and provided that $ d\mu_{\Delta, N,\vec \lambda}$ is chosen
in a class of probability measures such that this limit converges to
 a nonnegative invariant measure  $d\mu_{\Delta,\vec \lambda}$,
one obtains
\bea
\lim_{N \to \infty}
\int_{\mathcal S_N} e^{i\; \tr (\cJ U)}) d\mu_{\Delta, N,\vec \lambda} (U)
= \int_0^{\infty}  e^{- b\,\tr [\cJ^2]} d\mu_{\Delta,\vec \lambda}(b)\,.
\eea
Thus \eqref{lag} becomes,
\beq
L_{\infty,\vec\lambda}(\{\tr[\cJ^{2p}]\}_p)
= \lim_{\Delta \to 0} \Delta^{-1}\Big(\int_{0}^{\infty}
(1- e^{-b\, \tr (\cJ^2)}) d\mu_{\Delta, \vec \lambda} (b)\Big) \,,
\label{lag2}
\eeq
where $S_{\infty}$ is the set of symmetric matrices with infinite size.
The rest is very similar to the finite case. Decomposing $\int_{\mathcal S_\infty}$ as
$\int_{b> \Delta} + \int_{b\leq \Delta}$ and expand the
sector $b\leq \Delta$ as
\bea
 A^{\infty}_{\leq \Delta} &=&
\int_{b \leq \Delta}
(1- e^{-b\; \tr (\cJ^2)}) d\mu_{\Delta, \vec \lambda} (b) =
-\sum_{k=1}^{\infty} \frac{(-1)^{k}}{k!}[\tr (\cJ^2)]^{k} \int_{b\leq \Delta}
b^{k}\, d\mu_{\Delta, \vec \lambda} (b)\crcr
&=& [\tr (\cJ^2)] \int_{b\leq \Delta} b\, d\mu_{\Delta, \vec \lambda} (b)
 + \int_{b\leq \Delta} O(b^{2})\, d\mu_{\Delta, \vec \lambda} (b)\,,
\label{exps2}
\eea
where $O(b^2)$ includes all remaining terms in the expansion.
We perform a change of variables such that $b \to b'= b/\Delta$, and
obtain
\bea
A^{\infty}_{\leq \Delta} = \Delta [\tr (\cJ^2)] \int_{b'\leq 1} b'\, d\mu_{\Delta, \vec \lambda} (\Delta b')
 + \Delta^2 \int_{b'\leq 1} O(b'^{2})\, d\mu_{\Delta, \vec \lambda} (\Delta b')\,.
\label{exps3}
\eea
Re-injecting \eqref{exps3} in \eqref{lag2}, recalling that the limit \eqref{hyp}
holds, one can easily identify an $a$ and finally reach the result \eqref{larunimat}.
The condition \eqref{larmatcond} is equivalent to the convergence of the
term $\int_0^{\infty} (1- e^{-b\; \tr [\cJ^2]}) d\sigma_{\vec \lambda}(b) <\infty$.
This ends the proof of the theorem.

\qed

 Note that the order in which we have taken  the limits
$\lim_{\Delta \to 0}$ and $\lim_{N\to \infty}$ does not matter
for  obtaining $L_{\infty, \vec \lambda}$, namely
\beq
L_{\infty, \vec \lambda} = \lim_{\Delta\to 0} \lim_{N\to \infty} \mathcal{F}_{N,\Delta}=\lim_{N\to \infty}\lim_{\Delta\to 0}\mathcal{F}_{N,\Delta}= \lim_{N\to \infty}L_{N, \vec \lambda}\,,
\label{limlim}
\eeq
where $\mathcal{F}_{N,\Delta}=\Delta^{-1}\int_{\mathcal S_N}(1- e^{i\; \tr (\cJ U)}) d\mu_{\Delta, \vec \lambda} (U)$. This displays
the fact that the function $L_{\infty, \vec \lambda}$  is continuously connected to $L_{N, \vec \lambda}$.

\bigskip

\noindent{\bf Case of a non-symmetric matrix.}
Consider $M$ the initial matrix field now to be real but non symmetric.
The source function $J$ should be non symmetric and real as well.
In any case, we use a singular value decomposition for
matrix part of $J$ as $\cJ= V_1 \Sigma V_2^{\dag}$
where $\Sigma$ is a real nonnegative entry matrix, and where $V_{1,2}\in O(N)$. But $\Sigma$ contains in fact the non-negative square root of the eigenvalues of $\cJ^t \cJ$ or $\cJ \cJ^t$, that is
\bea
\tr (\Sigma^2) = \tr (\cJ^t \cJ)
\eea
which allows to perform all the above analysis. One recovers the analog of \eqref{unimat} for a generic matrix $\cJ$ as
 \beq
L_{N,\vec\lambda}(\{\tr [(\cJ^{t}\cJ)^p]\}_p)
 = a \,\tr [\cJ^t \cJ] + \int_{\tr (U^tU)>0} (1- e^{i\,\tr (\cJ^tU)}) d\sigma_{\vec \lambda}(U) \,,
\label{unimatgen}
\eeq
 where $a\geq 0$ and $d\sigma_{\vec \lambda}(U)$ is a non negative
invariant measure over the space of square matrices such that
\beq
\int_{\mathcal{M}_N} \tr(U^tU)/(1+\tr(U^tU)) d\sigma_{\vec\lambda}(U) < \infty\,.
\label{matcondgen}
\eeq
A similar analysis works for the case $N=\infty$. The formulation for complex
matrices can be inferred in the same manner.

\section{Applications: Finding the measure}
\label{sect:app}

Although we can address the case of non symmetric matrices,
for simplicity, we will perform the analysis for only
symmetric matrices.
The aim of this section is to provide an explicit formula
for the measures $d\sigma_{\vec\lambda}(U)$ and $d\sigma_{\vec\lambda}(b)$ such that
one may infer the other possible terms that one can introduce
in the initial model.

\subsection{Finite size matrices}
\label{subsect:appmN}

In the previous section, we have establish that, for
independent value matrix models,
\bea
Z(J,\vec\lambda) = e^{-\int L_{\vec \lambda}[\,\{\tr (J(x)^{2p})\}_{p}\,]dx}\,,
\eea
where $L_{\vec \lambda}[\,\{\tr (J(x)^{2p})\}_{p}\,]$ is of the most
general form given by the $O(N)$-invariant quantity
\beq
L_{N,\vec\lambda}(\{\tr [\cJ^{2p}(x)]\}_p)
 = a \, \tr [\cJ^2(x)] + \Big[\int_{\tr (U^2)>0} (1- e^{i\,\tr (\cJ(x) U)}) d\sigma_{\vec \lambda}(U)\Big] \,,
\label{unimat2}
\eeq
with $ d\sigma_{\vec \lambda}(U)$ an $O(N)$-invariant nonnegative measure over $\mathcal{S}_N$. In the remaining, the analysis will be restricted to the generic situation
such that $ d\sigma_{\vec \lambda}(U)=C^2(U)dU$.
Note that implicitly $C^2(U)$ should be a function of
the basic invariants $\{\tr [U^{2p}(x)]\}_p$ and $\vec \lambda$.
We are also interested only in the case $a=0$, hence in an interacting theory.

The use of $C(U)$ gives us a useful representation of the
field as follows. Let $A(x,U)$ and $A^\dag(x,U)$
be the annihilation and creation operators in the ordinary
sense such that they satisfy the commutation relation
\beq
[A(x,U),A^\dag(y,V)] = \delta_{\R^n}(x-y)\delta_{\mSn}(U-V)\,,
\label{comut}
\eeq
with a self-explanatory notation. We assume that there is a
vacuum state $|0\ket$ so that
\beq
A(x,U) |0\ket = 0 \,.
\eeq
Out of these initial operators, we built two new ones, namely
\beq
B(x,U)= A(x,U) + C(U)\,,  \qquad
B^\dag(x,U)= A^\dag(x,U) + C(U)\,,
\eeq
obeying the same relation \eqref{comut} and
$B(x,U)|0\ket = C(U)|0\ket$. From the $B$'s, we express the matrix field operator as
\bea
M(x) = \int_{\mSn} \,B^\dag(x,U) U B(x,U)\, dU \,,
\qquad
M_{ab}(x) = \int_{\mSn} \,B^\dag(x,U)\, U_{ab}\, B(x,U)\, dU \,.
\eea
As we will soon learn,
\bea
\Bra 0 \Big{|} e^{i\int \tr[J(x)\cdot M(x)] dx} \Big{|}0 \Ket
 = e^{-\int \Big[\int[1-e^{\tr[J(x)\cdot U]}]C^2(U)dU]\Big]dx}\,, \label{51}
\eea
and our task is to determine the connection between $C^2(U)$
and the model action functional. The bilinear representation of our basic operators means that
local products arise from an operator product expansion and not by Wick ordering. As a consequence, we have
\bea
M_{ab}(x)M_{cd}(y) &=& \int_{\mSn\times\mSn} \,B^\dag(x,U)B^\dag(y,V)\, U_{ab}V_{cd}\, B(x,U)B(y,V)\,  dUdV \crcr
&+&
\delta_{\R^n}(x-y)\int_{\mSn} B^\dag(x,U) U_{ab}U_{cd}\, B(y,U)\,  dU \,,
\crcr
&=& :M_{ab}(x)M_{cd}(y): + \delta_{\R^n}(x-y)\int_{\mSn} B^\dag(x,U) U_{ab}U_{cd}\, B(y,U)\,  dU \,.
\eea
We define ($R$ stands for ``renormalized'')
\beq
M^{2}_{R;abcd} =\beta\int_{\mSn} B^\dag(x,U) U_{ab}U_{cd}\, B(y,U)\,  dU \,,
\eeq
where $\beta$ has the dimension of $L^{-n}=$ dimension of $\delta_{\R^n}(0)$. For simplicity, hereafter, we choose the numerical value $\beta=1$.
Higher order products can be computed as well and, using the same prescription, will lead to $M^{w}_{R;a_1b_1\dots a_wb_w}$.

We are now in position to seek a relationship between $C^2(U)$ and the model.
Consider the action of a model given by \eqref{actmatrix}, $C^2(U)$ is such that
\bea
\Bra 0 \Big{|} e^{i\int \tr[J(x)\cdot M(x)] dx} \Big{|}0 \Ket
 = K\int e^{\int\Big[    i  \tr[J(x) M(x)] - \Big(\frac12\mu^2\,\tr[M^2(x)] +  V_{\vec \lambda}(\{\tr[M^{2p}(x)]\})\Big)\Big]dx   } dM
\label{mean}
\eea
with $K$ a constant normalization factor chosen so that the entire expression reduces to unity if $J\equiv0$.
Modifying the left hand side of the above equation, where $W$ is chosen so the left side is unity when $J\equiv0$,  we consider
\bea
W\,\Bra 0 \Big{|} e^{\int\Big[ i \tr[J(x)\cdot M(x)] dx -
\chi_{\Delta}(x) V'(\{\tr[M^{2p}(x)]\})\Big]dx} \Big{|}0 \Ket\,,
\eea
in which the interaction, in the right hand side  of \eqref{mean},
picks a term of the form
$\chi_{\Delta} V'(\{\tr[M^{2p}]\})$ and
$\chi_{\Delta}(x)$ keeps its previous meaning as an indicator function.
We can determine as well  how $C^2(U)$ gets modified under such a transformation. Using the $R$-product prescription for all
local products, one evaluates
\bea
&&
W\,\Bra 0 \Big{|} e^{\int\Big[ i \tr[J(x)\cdot M(x)] -  \chi_{\Delta}(x) V'(\{\tr[M_R^{2p}(x)]\})\Big]dx} \Big{|}0 \Ket
 =\crcr
&& W\,\Bra 0 \Big{|} e^{\int \Big\{ \int_{\mSn} \,B^\dag(x,U)
\big[ i\tr[J(x)U]  -  \chi_{\Delta}(x)V'(\{\tr[\prod_{i=1}^{2p}U_{a_ib_i}]\}) \big]B(x,U)\,  dU\Big\} dx} \Big{|}0 \Ket \crcr
&&= W\,\Bra 0 \Big{|}  \Big(\, e^{\int \Big\{ \int_{\mSn} \,B^\dag(x,U)
\Big[ i\tr[J(x)U]  -  \chi_{\Delta}(x)V'(\{\tr[\prod_{i=1}^{2p}U_{a_ib_i}]\}) \Big]B(x,U)\,  dU\Big\} dx} \, \Big)_R\,\Big{|}0 \Ket \label{eqi}\\\cr
&&
= \frac{e^{-\int dx \big[ \int \big(
1- e^{i \tr [J(x)\cdot U]  -  \chi_{\Delta} V'(\{\tr[U^{2p}] \}) } \big) C^2(U)dU \big] } }{e^{-\int dx \Big[ \int \big(
1- e^{ -  \chi_{\Delta} V'(\{\tr[U^{2p}] \}) } \big) C^2(U)dU \Big] }}\,;
\label{ratio}
\eea
note the change in the line \eqref{eqi} for $R$-product
which gives sense to the  local products.
To obtain the final form of this equation, there is an intermediate step in the above calculation which can be explained by a simple exercise using canonical (for the harmonic
oscillator for instance) coherent states expectation values
(in the usual notation such that $[a,a^\dag]=1$, $b_n=1/\sqrt{n!}$)
\beq
\langle z|e^{i c a^\dag a }|z\rangle
 =\langle z|e^{ic}z\rangle =  e^{-|z|^2}
\sum_{n,m} \langle n| b_n\bar z^{n} b_m (e^{ic}z)^m|m\rangle
=
e^{z^*(e^{ic}-1)z} = \langle z|:e^{a^\dag (e^{ic}-1)a}:|z\rangle \,,
\eeq
using the fact that $a|z\rangle=z|z\rangle$;  incidentally, an extension of this argument may be used to verify \eqref{51}.
Now, coming back to \eqref{ratio},
we let $\Delta_*$ expand  to cover all $\R^n$, and thus
\bea
&&
\lim_{\Delta\to \infty} W\, e^{-\int dx \Big[ \int \Big(
1- e^{i \tr [J(x)\cdot U]  -  \chi_{\Delta} V'(\{\tr[U^{2p}] \}) } \Big) C^2(U)dU \Big] } = \crcr
&& \lim_{\Delta\to \infty} W\,e^{-\int dx \Big[ \int \Big(
e^{ \chi_{\Delta} V'(\{\tr[U^{2p}] \}) } - e^{i \tr [J(x)\cdot U]} \Big)e^{-  V'(\{\tr[U^{2p}] \}) }  C^2(U)dU \Big] } \crcr
&&=
e^{-\int dx \Big[ \int \Big(
1 - e^{i \tr [J(x)\cdot U]} \Big)e^{-  V'(\{\tr[U^{2p}] \}) }  C^2(U)dU \Big] }
\eea
for the right normalization factor $W$, which is simply the denominator in \eqref{ratio}. Therefore, we find
the effect on the measure on the matrix space given by
\bea
S \to S + \int  V'(\{\tr[M^{2p}(x)]\}) dx \,,
\qquad
C^2(U) \to e^{- V'(\{\tr[U^{2p}]\}) } C^{2}(U)\,.
\eea
As a special choice, the potential $V'(\{\tr[M^{2p}(x)]\}) = -V(\{\tr[M^{2p}(x)]\})$, simply {\it cancels}
the original nonlinear interaction leaving only the mass term such that
\bea
Z_{PF}(J,\mu) = K\int e^{\int\big[    i  \tr[J(x)\cdot M(x)] -\frac12\mu^2\, \tr[M_{R}^2(x)]\big]dx   } dM\,.
\label{zpf}
\eea
As expected, this is not the characteristic functional of the free-theory because of the $R$-multiplication prescription on the ``quadratic term''
$\tr[M_{R}^2(x)]$. Instead, \eqref{zpf} represents the {\it pseudo-free matrix theory}, i.e., the model continuously connected to the interacting theories.
In order to characterize the pseudo-free model, let us investigate
a special change in the matrix measure. From the same procedure used above, we obtain
\bea
Z_{PF}(J,\mu) = e^{-\int \Big[ \{1-e^{\tr[J(x) U]}\}e^{-\frac 12\mu^2 \tr[U^2]}C_0^2(U)dU     \Big]dx}\,.
\label{zpf2}
\eea
Note that, for any constant $\alpha$, from \eqref{zpf}
one learns that
\beq
Z_{PF}(\alpha J,\alpha\mu) =  Z_{PF}(J,\mu)\,.
\eeq
Performing the same scale transformation, using now \eqref{zpf2} leads to
\bea
Z_{PF}(\alpha J, \alpha\mu) = e^{-\int \Big[ \{1-e^{\tr[J(x) (\alpha U)]}\}e^{-\frac 12\mu^2\tr[(\alpha U)^2]}C_{0}^2(\alpha U)\,d(\alpha U)\,\Big]dx}
\eea
which can be scale invariant if and only  if
\beq
C_0^{2}(\alpha U)\, d(\alpha\,U) = C_0^2(U)\,dU\,, \qquad {\rm hence} \qquad C_0^2(\alpha)=\Gamma\,|\alpha|^{-N^*}\,,
\label{keep}
\eeq
for some constant $\Gamma$ and, as before, $N^*=N(N+1)/2$.
One should keep in mind that we assume that there exists  a positive integer $p_{\max}$
such that $C_0^2(U) = C_0^2(\{\tr[U^{2p}]\}_{p=1}^{p_{\max}})$. At this point, \eqref{keep} implies that
\bea
C_0^2(\frac{U}{\alpha}) = C_0^2(\{\tr[\frac{U^{2p}}{\alpha^{2p}}]\}_p)\,.
\label{cco}
\eea
 Given an even integer $q \in \mathbb{N}$, we introduce the set $\mathfrak{P}_q$ of nontrivial partitions $q$ and the set $\mathfrak{P}_q^*$ of nontrivial partition of $N^*+q$, such that
\bea
&&
\cP_{q,I} \in \mathfrak{P}_q \,,\qquad 1 \leq I \leq q\,, \qquad
\cP_{q,I} = (N_1, \dots, N_I) \,, \qquad
\crcr
&&
q= \sum_{i \in I} 2p_iN_i\,, \qquad 1\leq N_i\leq q\,,
\quad 1\leq p_i \leq p_{\max}\,,
\cr\cr
&&
\cP^*_{q,I} \in \mathfrak{P}_q^* \,,\qquad 1 \leq I \leq N^*+q\,, \qquad
\cP^*_{q,I} = (N_1, \dots, N_I) \,, \qquad
\crcr
&&
N^* +q= \sum_{i \in I} 2p_iN_i\,, \qquad 1\leq N_i\leq N^*+q\,,
\quad 1\leq p_i \leq p_{\max}\,.
\eea
Then, one writes a general solution for \eqref{cco} as
\beq
C_0^2(\{\tr[U^{2p}]\}_p) =
\sum_{q=0;q \,\text{even }}^{q_{\max}}
\sum_{\mathfrak{A}_{q}\subset\mathfrak{P}_q\,,\,\mathfrak{A}^*_{q}\subset\mathfrak{P}^*_q}
g_{(\mathfrak{A}_{q};\mathfrak{A}^*_{q})}  \frac{\sum_{\cP_{q,I}\in\mathfrak{A}_{q}}
 g^{(1)}_{\{N_i\}_{I}} \prod_{i\in I} (\tr[U^{2p_i}])^{N_i}}{\sum_{\cP^*_{q,I'}\in \mathfrak{A}^*_{q}}
g^{(2)}_{\{N_j\}_{I'}} \prod_{j\in I'} (\tr[U^{2p_j}])^{N_j}}
\label{gener}
\eeq
where $q_{\max}$ is an arbitrary finite even integer
(the case $q_{\max}=\infty$ might lead to convergence issues
that we shall avoid), $g_{(\mathfrak{A}_{q};\mathfrak{A}^*_{q})} $
 are positive constants which contain some
dimensional normalization coming from the $R$-regularization
of the two-point function, $g^{(1,2)}_{\{N_i\}}$ are
also constants which can be chosen without dimension;
the  sum $\sum_{\mathfrak{A}_{q}\subset\mathfrak{P}_q\,,\,\mathfrak{A}^*_{q}\subset\mathfrak{P}^*_q} $ is performed
over all subsets $\mathfrak{A}_{q}$ of $\mathfrak{P}_q$ and
$\mathfrak{A}_{q}^*$ of $\mathfrak{P}^*_q$;
in the ratios, the sums are performed over elements of these
subsets $\mathfrak{A}_{q}$  and
$\mathfrak{A}_{q}^*$ consisting in partitions themselves.
 Note that this solution may be not the most general one nor
is it unique. However, it provides a wide class of solutions materializing the fact that  each ratio in \eqref{gener} should scale as $\alpha^{N^*}$ after mapping
$U \to U/\alpha$.  As an illustration, this solution includes the
following kind of terms, assuming that $N^*$ is large enough,
\bea
&&
 \frac{1}{(\tr[U^{2}])^{\frac{N^*}{2}}}\,, \quad
\frac{1}{(\tr[U^{4}])^{\frac{N^*}{4}}}\,, \quad  \frac{1}{(\tr[U^{2}])^{N_1}(\tr[U^{4}])^{\frac{(N^*-2N_1)}{4}}}\,,
\quad \frac{1}{(\tr[U^{2}])^{N_1}(\tr[U^{6}])^{\frac{(N^*-2N_1)}{6}}},
\cr\cr
&&  \\
&&
\frac{1}{g^{(2)}_{1}(\tr[U^{2}])^{\frac{N^*}{2}}
+ g^{(2)}_{2} (\tr[U^{6}])^{\frac{N^*}{6}}} \,,
\quad \frac{ g^{(1)}_{1}(\tr[U^{2}])^{\frac{q}{2}}
+ g^{(1)}_2(\tr[U^{4}])^{\frac{q}{4}}}{g^{(2)}_{1}(\tr[U^{2}])^{\frac{N^*+q}{2}}
+ g^{(2)}_{2} (\tr[U^{2}])^{N_1}(\tr[U^{6}])^{\frac{(N^*+q-2N_1)}{6}}}.
\nonumber
\eea
In summary, the functional integral of an
$O(N)$-invariant matrix model is given by
\bea
Z(J,\vec \lambda) = K \int e^{i \int \tr[J(x) M(x)]dx - \int \big\{\frac12\mu^2 \tr[M^2(x)] + V_{\vec \lambda}(\{\tr[M^{2p}(x)]\}_p)\big\}dx }.
\label{zp}
\eea
When the theory is genuinely interacting, we have $\vec\lambda\neq 0$,
and \eqref{zp} reduces to
\bea
Z(J,\vec \lambda)  = e^{-\int \Big\{ \int  [1-e^{\tr[J(x)U]}]
e^{-\frac12 \mu^2\tr[U^2] - V_{\vec\lambda}(\{\tr[U^{2p}]\}_p)}     C_0^2(\{\tr[U^{2p}]\}_p) dU\Big\}dx}\,.
\label{true}
\eea
In the limit that the coupling constants all vanish, $\vec \lambda=0$, this generating functional
yields the pseudo-free theory,
\bea
Z_{PF}(J,\mu)  = e^{-\int \Big\{ \int [1-e^{\tr[J(x)U]}]
e^{-\frac12 \mu^2 \tr[U^2]} C_0^2(\{\tr[U^{2p}]\}_p) dU\Big\}dx}\,,
\eea
which differs significantly from the generating functional for the free theory,
\bea
Z_F(J,\mu) =
e^{- \frac{1}{2\mu^2} \int \tr[J^2(x)]dx}
\label{zfree}
\eea
obtained from \eqref{zp} by formally putting $\vec\lambda=0$ and
computing the remaining functional integral as a traditional Gaussian functional integral.
The interacting theory provides a continuous perturbation
of the pseudo-free theory and a discontinuous perturbation
of the free theory. This also shows that the results for the $O(N)$-invariant, finite-component, {\it vector} case \cite{Klauder:1975kd} basically extend to the $O(N)$-invariant  (symmetric) {\it matrix} models with finite size. Since we did not use in the above calculation any feature about the symmetric property of the matrices, one can reasonably infer that similar results hold for general square matrices, according to the discussion in Section \ref{sect:mat} in the paragraph: Case of a non-symmetric matrix.

\subsection{Infinite size matrices}
\label{subsect:inf}

The task now is to determine the characteristic functional of an infinite
size independent-value matrix model from a limit of those with finite size
(we recall that the limits defining $L_{\infty,\vec \lambda}$ commute, i.e.,
specifically, \eqref{limlim} holds).
For this limit to hold,  one must pay attention to the parameters
$V_N$, $\mu_N$ and the family of arbitrary couplings $g_{(\cdot;\cdot)}, g^{(1)}_{(\cdot)}$, and $g^{(2)}_{(\cdot)}$.

In the finite case, we start with the solution \eqref{true} above for $L_{N,\vec \lambda}(\{\tr[J^{2p}(x)]\})$ in the interacting theory (i.e., $a=0$ in \eqref{unimat2}). Let us again restrict to a small sector $\Delta_* $ of $\R^n$ and consider the part of that functional independent of the positions as
\beq
L_{N,\vec \lambda}(\{\tr[\cJ^{2p}]\}_p) =
\int  [1-e^{i\tr[\cJ U]}] e^{-\frac12 \mu_N^2\tr[U^2] - V_{N,\vec\lambda}(\{\tr[U^{2p}]\}_p)}     C_{0,N}^2(\{\tr[U^{2p}]\}_p) dU\,.
\eeq
Due to the intricate expansion of the solution
$C_{0,N}^2(\{\tr[U^{2p}]\}_p)$ in terms of several matrix invariants, the $N=\infty$ limit becomes difficult to track.

To proceed further, it is useful to restrict attention to a more narrow range of models. Thus, let us  restrict the analysis to a
potential of the form
\beq
V_{N,\vec\lambda}  (\{\tr[U^{2p}]\}_p) =
 V_{N,\lambda}(\tr[U^{2}])\,.
\eeq
Also some limitation on the general solution for $C_{0,N}^2(U)$ is in order, such as
\beq
  C_{0,N}^2(\{\tr[U^{2p}]\}_p) =
  C_{0,N}^2(\tr[U^{2}]) = g_{N}\frac{1}{(\tr[U^2])^{\frac{N^*}{2}}}\;,
\eeq
 which is not unreasonable since within the functional integral expressions of the form $[\tr(U^2)]^p$
 dominate, and do so greatly for large $N$, over homogeneous expressions such as $[\,\tr(U^{2p})\,]$.
In particular, let us consider
\bea
L_{N, \lambda}(\tr[\cJ^{2}]) = g_N
\int  [1-e^{i\tr[\cJ U]}] e^{-\frac12 \mu_N^2\tr[U^2] - V_{N,\lambda}(\tr[U^{2}])}    \frac{1}{(\tr[U^2])^{\frac{N^*}{2}}}dU\,.
\eea
Following similar procedures to those in Section \ref{sect:mat}, we
proceed by diagonalizing $\cJ= O j O^t$ and introducing $r^2=\tr[U^2]$ and spherical coordinates \eqref{diagonal}, such that
\bea
\lim_{N \to \infty} L_{N, \lambda}(\tr[\cJ^{2}]) &=&
\lim_{N \to \infty} g_N
\int  \Bigg\{ [1-e^{i r\, \sum_{a}j_{aa} \Big(\prod_{l=1}^{a-1}\sin\theta_l\Big) \cos\theta_{a}} ]\Big[\prod_{l=1}^{N} (\sin\theta_l)^{N^*-(l+1)}d\theta_l \Big]\crcr
&&  e^{-\frac12 \mu_N^2 r^2 - V_{N,\lambda}(r^2)} \frac{dr}{r}    d\Omega_{N^*-(N+2)}
  \Bigg\} .
\eea
Using  steepest descent techniques for integrating $\theta$ as done in Section \ref{sect:mat}, introducing
\beq
r\to \sqrt{2bN^*}\,, \qquad \mu_N = \frac{1}{N^*}\mu \,, \qquad
V_{N,\lambda}(2bN^*) = V_{\lambda}(b)\,,
\eeq
 and using $g_N$ to neutralize the contributions depending on
$N$ coming from the integration and the change of variables,
we have
\bea
L_{\infty,\lambda} (\tr[\cJ^{2}])
= g \int_{0}^{\infty}  [1-e^{- b \,\tr[\cJ^2] } ]
 e^{-\frac12 \mu^2 b - V_{\lambda}(b)} \,\frac{db}{b}    \,.
\eea
Setting $\lambda\to 0$ in this expression yields the pseudo-free
theory and, now returning to the full position space, one obtains
\bea
&&-\ln[\,Z_{PF}(J)\,] =\int L_{PF} (\tr[J^{2}])\,dx
= g \int\Big\{\int_{0}^{\infty}  [1-e^{- b \,\tr[J^2(x)] } ]
 e^{-\frac12 \mu^2 b} \,\frac{db}{b} \Big\} dx\nonumber\\
  &&\hskip6.29em= g \int \ln\big[\,1 +\frac{2}{\mu^2}\tr[J^2(x)]\big] dx \,,
\label{lpf}
\eea
which should be compared with the free theory
\bea
-\ln[\, Z_F(J)\,]=\int L_{F} (\tr[J^{2}(x)]\,dx)
= \int  \frac{2}{\mu^2}\tr[J^2(x)]\, dx \,.
\eea

 It is interesting to observe that if the factor $g=1$ then for weak values of the source $J(x)$ and/or large values of the mass parameter $\mu$---or more precisely, $\tr[J^2(x)]/\mu^2\ll1$---the pseudo-free theory essentially agrees with the free theory. It may be argued that this fact can be used to fix the value of $g$. However, the functional form of the pseudo-free and free theories remains manifestly different over the entire range of the strength of the source, which includes both large values as well as small values. Thus one should not read too much into the similarity of the behavior for a limited range of parameters. For example, any discussion of ``perturbation'' about some form of a ``free theory'' would necessarily involve large and small source values and would therefore explore the fundamental differences between the pseudo-free and free theories.

Alternatively, one notices that all previous developments leading to
\eqref{lpf} can be reached
using the field definition $M(x) = \int_{\mathcal{S}_{\infty}} \,B^\dag(x,U) U B(x,U)\, d\rho(U) $ for infinite size matrices. This makes sense if one introduces a weight function $\rho(U)$ so that the measure $d\rho(U)$ is well defined on an infinite
dimensional space. The operator $B$ is defined as $B(x,U) = A(x,U) + 1$ with
$[A(x,U),A^\dag(y,V)] = \delta_{\R^n}(x-y)\delta_{\mSi}(U-V)$,
where $\delta_{\mSi}(U-V)$ is understood in the distribution
sense with respect to the measure $d\rho(U)$, namely
\beq
\int \delta_{\mSi}(U-V) d\rho(U) =1\,.
\eeq
The notion of $R$-ordering extends in the present
setting as well. The rest of the analysis  naturally follows and leads to
\eqref{lpf}. The same conclusion can be inferred: the interacting theory
is continuous connected to the pseudo-free theory \eqref{lpf}.

\section{Conclusion}
\label{sect:concl}

We have investigated independent-value matrix models and extended
results obtained in the vector situation \cite{Klauder:1975kd,Klauder:1975kc} to both finite and infinite size matrices. We first find the general
formula of a functional over matrix invariants in terms of an invariant measure.
The determination of that measure amounts to specify invariant actions.
As another interesting result in such invariant models, we find that the interacting theory is again continuously connected with the so-called pseudo-free theory and not with the free theory. This has been proved for finite size $N$ matrix models with an $O(N)$ invariant action which incorporates invariants of any order $\{\tr[M^{2p}]\}_p$.
The case $N=\infty$ is more peculiar but we succeed to prove the similar result for $O(\infty)$ invariant matrix models equipped with an interaction as a general function of a unique basic invariant $\tr[M^2]$. A way perhaps to extend our developments
to other types of matrix invariants such as at least $\tr[M^4]$ would be to apply resolvent  methods as used in the framework of statistical mechanics in random
matrices models \cite{difrance}. This aspect deserves to be analyzed.

This study is part of a larger program   which advocates that the expansion in perturbation theory should not be performed around the free theory but the one that is continuously connected to the interacting theories, namely the pseudo-free theory, when that theory differs from the free theory.
Concerning this, the vector case has been resolved, and the matrix
case, although not totally resolved, seems to support similar conclusions according to the present work. Then one may naturally ask how the present formalism might be led further by discussing the case of independent-value, multi-index, tensor models. We
point out that, recently, a basis of unitary invariants has been highlighted in the framework of colored tensor models \cite{color,Gurau:2011kk,Gurau:2012vk,Gurau:2013pca}.
Tensor invariants can be traced back for years \cite{gord} but have been
rediscovered after analyzing of the $1/N$ expansion of colored random tensors \cite{Gurau:2012vk}. Their Gaussianity at large $N$ proves to be universal \cite{Gurau:2011kk,Gurau:2012vk}. We expect that, using a different method
by introducing a position space $x$ attached to these tensors, and several ingredients of this work, we might find in the tensor situation another behavior
at large $N$ when the theory is fully in interaction.

\section*{Acknowledgements}
JRK thanks the Perimeter Institute, Waterloo, Canada, for its hospitality.
Research at Perimeter Institute is supported by the Government of Canada through
Industry Canada and by the Province of Ontario through the Ministry of Research and Innovation.

\end{document}